\documentclass[traditabstract]{aa}
\usepackage{graphicx}
 \usepackage{lscape}
 \usepackage{colortbl}
 \usepackage{natbib}
 \usepackage{verbatim} 
  \usepackage{comment} 
\usepackage{ulem}
\usepackage{url}
 \bibpunct{(}{)}{;}{a}{}{,}
\usepackage{txfonts}
\begin{document}
\input psfig.tex
\def\simgt{\stackrel{>}{{}_\sim}}
\def\simlt{\stackrel{<}{{}_\sim}}

\titlerunning{A low-scatter survey-based mass proxy}

\title{A low-scatter survey-based mass proxy for clusters of galaxies}
\author{S. Andreon} 
\institute{INAF-Osservatorio Astronomico di Brera, via Brera 28, 20121, Milano, Italy\\
\email{stefano.andreon@brera.inaf.it} 
}
\date{Received --, 2012; accepted --, 2012}

\abstract{
Estimates of cosmological parameters using galaxy clusters have 
the scatter in the observable at a given mass as a
fundamental parameter. 
This work computes the amplitude of the 
scatter for a newly introduced mass proxy, the
product of the cluster total luminosity times the mass-to-light
ratio, usually referred as stellar mass.
The analysis of 12 galaxy clusters with excellent total
masses shows 
a tight correlation between the stellar mass, or stellar
fraction, and total mass within $r_{500}$ with negligible
intrinsic scatter:  the 90\% upper limit is $0.06$
dex, the posterior mean is $0.027$ dex.  This scatter is similar to the one
of best-determined mass proxies, such as $Y_X$, i.e. the
product of X-ray temperature and gas mass. 
The size of the cluster sample used to determine
the intrinsic
scatter is small, as in previous works proposing low-scatter proxies 
because very accurate masses
are needed to infer very small values of intrinsic scatter.
Three-quarters of 
the studied clusters have $lgM\la 14$ $M_\odot$, which is advantageous
from a cosmological perspective because these clusters are far more
abundant than more massive clusters.
At the difference of other mass proxies such as $Y_X$, 
stellar mass can be determined with survey data 
up to at least $z=0.9$ using
upcoming optical near-infrared surveys, such as DES and Euclid,
or even with currently available surveys, covering however 
smaller solid angles. On the other end, the uncertainty about
the predicted mass of a single 
cluster is large, $0.21$ to $0.32$ dex, depending
on cluster richness. This is largely because the proxy itself has
$\approx 0.10$ dex errors for clusters of $lgM\la 14$ $M_\odot$ mass. 
}

   \maketitle

\section{Introduction}

Galaxy clusters are the most massive bound structures.
These structures, as well galaxy groups, 
in this work are called halos, a term that does not require an arbitrary
boundary between the two types of objects.
Halos are both cosmological probes and giant physics laboratories.
Halos are multicomponent systems formed mostly by dark matter 
and by baryons in the form of gas and stars. The proportion of gas (intra-cluster
medium) and star components is
10:1 at $lgM \sim 14.5$ $M_\odot$, but 1:1 at $lgM \sim 13.5$ $M_\odot$  
(e.g. Andreon 2010 and reference therein) and 1:2 at $lgM \sim 12.5-13$ $M_\odot$
(Dai et al. 2010; Humphrey et al. 2011, 2012). The cluster mass function is
steep (e.g. Press \& Schecther 1974; Jenkins et al. 2001), 
and therefore most halos have low 
masses and thus comparable amounts of stars and gas. 
Therefore, in all but the shallowest (in mass) surveys, galaxies and
gas contain similar amounts of baryons. 

The efficient use of clusters as cosmological probes requires a 
low-scatter mass proxy because cosmological constraints become looser and looser
with increasing scatter between mass and mass proxy (e.g. Lima \& Hu 2005).
Quantities derived from survey data have generally large
scatter: richness and not-core-excised X-ray luminosities
have a $0.19$ dex intrinsic scatter (richness: Andreon \& Hurn 2010, 
X-ray luminosity: Vikhlinin et al.
2009a, Mantz et al. 2010a, see Andreon \& Hurn 2010 for
a comparison of the predicted mass uncertainties of these two proxies). Instead, 
the intrinsic scatter of the SZ mass proxy 
has yet to be robustly measured from the data (Allen, Evrard, \& Mantz 2011).
Proxies based on higher quality data perform better: core-excised 
X-ray luminosities have a small, and for this reason poorly determined, 
intrinsic scatter of 0.015-0.025 dex (Mantz et al. 2010a) above
$lgM \sim 14.5$ $M_\odot$ and an unknown scatter below; $Y_X$ shows
a similarly small and poorly determined intrinsic scatter of $0.052$ dex  above
$lgM \sim 14.5$ $M_\odot$ (Mantz et al. 2010a) or $0.07$
above $lgM \sim 14$ $M_\odot$ (Arnaud et al. 2007). 

More complex proxies, which combine information from, say, X-ray
and SZ data, should perform even better than those using only one
of them. In practice, one should have high-quality 
data for both proxies, not just for one of them. This usually strongly
reduces the usable sample and makes it harder to derive the selection function 
needed to determine of cosmological constraints.

The scatter in observable at a given mass, and not the other way
around, is a parameter appearing explicitly in the 
estimates of cosmological parameters based on galaxy
clusters counts and is thus a fundamental parameter. In fact, 
$\sigma_{proxy|M}$ has
been used in past cosmological estimates using cluster counts (Vikhlinin et
al. 2009, Mantz et al. 2010b, Rozo et al. 2010). To understand why
this scatter, and not the other way around, appears in cosmological estimates 
one must recall that
cosmological parameters are constrained by
changing them until the predicted abundance, as a function of the
observable, matches the observed distribution. Thus, one needs to go 
from the mass function, $p(M)$, to the predicted distribution in the observable, 
$p(proxy)$. Since $p(proxy) \propto p(proxy|M)*p(M)$
(Bayes' theorem, see also Lima \& Hu 2005 and Rozo et al. 2010 for its 
cosmological application), $p(proxy|M)$  (whose
minimal parametrisation is given by the scatter at
a given mass, $\sigma_{proxy|M}$) is needed.  One of the aims
of this work is to compute this quantity and compare it to
the scatter of other mass proxies, completing the similar exercise 
done in the review paper of Allen et al. (2011) for other mass proxies
(their Sect. 3.3.4) 

It is worth emphasizing that the scatter computed the other way around,
in mass at a given observable,
is also an interesting quantity because it is an approximation of 
how well one may estimate the mass of a {\it single} cluster. To be precise,
$\sigma_{M|proxy}$ alone is not
enough. The whole probability distribution $p(M|proxy)$ is needed\footnote{ 
An appreciation of the difference 
between $p(x|y)$ and $p(y|x)$ may, following D'Agostini (2012),  
be obtained by replacing $x$ with $woman$ and $y$
with $senator$, obtaining $\approx 10-60$\% for the former (in most countries) 
and $\ll 0.1$\% for the latter.} because one should also at least account  
for the proxy uncertainty and for the uncertainties in the
mean relation between mass and proxy. The width of the posterior
predictive distribution encapsulates all these uncertainties (including 
less obvious ones, such as errors coming from extrapolations)
and has been used in  Andreon \& Hurn (2010) to compare the performances of $L_X$ and 
$n_{200}$ as estimators of the mass of individual clusters. 

We emphasise that, once the intrinsic scatter is smaller than the mass error,
its value is poorly determined (for all mass proxies) and should be
read as an upper limit because of the uncertainty
of the mass error and of the likelihood function (often assumed to be Gaussian). 
Small changes of them imply large changes in the derived intrinsic scatter
when the latter is comparable to mass errors. 
So,
although in principle an intrinsic scatter much smaller than the
mass error can be
determined from data of arbitrary precision, in practice values smaller than 
mass errors should read
as upper limits.

In this paper, we introduce a new mass proxy, the stellar mass,
which shows a negligible small intrinsic scatter with hydrostatic
mass. Stellar mass is derived as in previous works
(e.g. Gonzalez, Zaritsky, \& Zabludoff 2007; Andreon 2010): it is given by 
the stellar M/L ratio times the total cluster luminosity. Three-quarters of 
the studied clusters have $lgM\la 14$ $M_\odot$, which is advantageous
from a cosmological perspective because these clusters are far more
abundant than more massive clusters. Furthermore, we show that stellar
masses can be derived from imaging data, making it possible to determine the
mass proxy up to at least $z=0.9$ with both current and 
upcoming imaging surveys.

An optical mass proxy with small scatter is of paramount 
importance, given the current large investment in optical and 
near-infrared surveys, such as 
the Dark Energy Survey (DES\footnote{http://www.darkenergysurvey.org/})
(Abbott et al. 2005), Euclid\footnote{http://www.euclid-ec.org} 
(Laureijs et al. 2011), and the planned 
Large Sinoptic Survey Telescope (Ivezic et al. 2008). All these
surveys cover
thousands of square degrees in several bands with a depth appropriate
to derive the proxy value to $z=1$ and beyond (see Sect.~3.3 for details).

Throughout this paper, we assume $\Omega_M=0.3$, $\Omega_\Lambda=0.7$, 
and $H_0=70$ km s$^{-1}$ Mpc$^{-1}$. Magnitudes are quoted in their
native system (quasi-AB for Sloan Digital Sky Survey (SDSS) magnitudes).  
All logarithms in this work are on base ten.

\section{Data and sample}

We started from clusters in Vikhlinin et al. (2006) and
Sun et al. (2009) with: a) hydrostatic mass errors lower than 25\% to retain
a sample of clusters with accurately measured masses. In practice,
because of the other constraints, the typical mass error of halos
retained in the sample is  0.05 dex; b) inside the
SDSS $8^{th}$ data release (Aihara et al. 2011); c) with redshift between
$0.02<z<0.14$, and d) mass larger than $10^{13.2}$ $M_\odot$ to keep
clusters for which the stellar mass can be accurately measured.
In practice, the redshift range of the retained
sample, listed in Table 1, is narrower than imposed ($0.02<z<0.078$) 
because of the other constraints. 
The retained sample consists of 12 unique clusters plus two repeats,
i.e. objects listed in both catalogs (MKW4 and Abell 1991) and 
analysed twice to check analysis-dependent systematics. 

The studied sample is based on
clusters with available masses, and we have not filtered it by
using any optical cluster
property, such as the presence of a prominent red sequence or a 
centrally located brightest cluster galaxy. Therefore,
while we are sure that we have not introduced
a selection bias, both the starting and the retained samples 
have an unknown representativity. Nevertheless, the starting
sample should not be completely unrepresentative, because it has been 
successfully used to calibrate the mass-$Y_X$ 
scaling relation for cosmological estimates (e.g. Vikhlinin et al. 
2009).

Measurements are performed within the $r_{500}$ 
radius\footnote{$r_{\Delta}$ is the radius within which the enclosed average
mass density is $\Delta$ times the critical density.}, 
listed in Vikhlinin et al. (2006) and Sun et al. (2009), and within
the virial radius, $r_{200}$, derived from
$r_{500}$ and $c_{500}$ (also listed in these papers). 
For three clusters (3C442, A2092, and NGC6269), the $r_{200}$ computation
assumes a concentration index of four, lacking $c_{500}$
for these two clusters in these papers.

The analysis of the optical data strictly follows Andreon (2010, 
A10 hereafter).
The basic data used in our analysis are $g$ and $r$ photometry
from SDSS, down to $r=19$ mag. The latter is the value where
the star/galaxy separation becomes uncertain (e.g. Lupton et al. 2002)
and is much brighter than the SDSS completeness 
limit (e.g. Ivezi{\'c} et al. 2002).  Specifically,
we use ``cmodel" magnitudes for ``total" magnitudes 
and ``model" magnitudes  for colours. 
We computed the mass in stars from the cluster total luminosity,
the latter computed as the integral of $L$ times the
luminosity function of red galaxies. We adopted a Bayesian approach,
as done for other clusters (e.g. Andreon 2006, 2010, Andreon et al. 
2006,  2008b, Meyers et al. 2012), and we account for the background
(galaxies in the cluster line-of-sight), which is estimated  
outside the cluster turnaround radius. We adopted
a Schechter (1976) luminosity function and
the likelihood expression given in Andreon, Punzi \& Grado (2005), which
is an
extension of the Sandage, Tammann \& Yahil (1979) likelihood expression  
for the case where a background is present.  The Bayesian approach allows us
to easily propagate
uncertainties and their covariance into the integral of the 
luminosity function and into the stellar mass. For example, in contrast to
all other authors (except Meyers et al. 2012), we do
not assume to complete knowledge of the faint end slope of luminosity function
and instead marginalise over it, allowing us to propagate
this uncertainty in derived quantities. As a visual check, we also
computed the luminosity function by binning galaxies in magnitude bins 
(e.g. Zwicky 1957, Oemler 1974, and many papers since then).

In the luminosity function computation, attention should be paid to
saturated stars misclassified as galaxies (we inspected the SDSS images
and removed them by hand) and 
to the brightest cluster galaxy (BCG), which
might not be drawn from the Schechter (1976) function 
(e.g. Tremaine \& Richstone 1977). To deal the latest issue, we
fit all galaxies, excluding the BCG, and added its luminosity 
contribution separately.  Finally, we removed by hand
very bright (much brighter and larger
than the BCG) unrelated galaxies that were spectroscopically
in the fore/background of the cluster. 

In this paper, we define as red galaxies those 
within 0.1 redward and 0.2 blueward in $g-r$ of the colour--magnitude 
relation, as in Andreon \& Hurn
(2010) and A10, and in agreement with works mentioned there.

Because of the shallow nature of the SDSS, it misses the 
flux coming from the galaxy outer regions (e.g. Blanton et al. 
2001, Andreon 2002). For this reason we corrected the measured flux by 15\% 
(Blanton et al. 2001).
To compute the luminosity in a sphere from the measured values (derived
in a cylinder), we assumed a Navarro, Frank \& White (1997) profile.
Finally, luminosities are converted in stellar mass by
adopting the $M/L$ value derived by Cappellari et al. (2006). 

To check whether our stellar mass are over/underestimated,  we
performed an approximate calculation using simulations, that account
only for the effect of the finite sample size of cluster galaxies,
and of the background subtraction 
because these two terms were found to be the main sources of error. 
First, we simulated 1000 clusters, all  composed of
73 galaxies whose luminosities are drawn from a Schecther (1976)
function with slope $\alpha=-1$. These are best--fit values of
NGC6269. As for real NGC6269 data, luminosities are drawn down to
$M^*+5$. To obtain the total luminosity of each simulated cluster, we
summed luminosities of the simulated galaxies (Schechter's draws). Now,
we needed to simulate the uncertainty associated to the background
subtraction.  We took a power-law to describe the background counts
with best--fit parameters measured for NGC6269
background area (slope $0.4$ and 27 background galaxies per unit
cluster solid angle). Next, we added the luminosity of a
realization of the background and removed the luminosity of another
realization to get the total net cluster luminosity. Since the number
of background galaxies is subject to Poisson fluctuations,  the
number of draws for the two background realizations is allowed to
fluctuate Poissonianly. Finally, since the background for NGC6269 is
estimated on a solid angle 12 times larger than the cluster area, 
our simulation also modelled this.  The total net luminosity showed a
scatter, from simulation to simulation,  of $0.11$ dex. For NGC6269
we derived $0.12$ dex from real data (Table 2), the latter also
accounting for other (minor) sources of error  not modelled in this
simulation. The two estimates are close enough not to warrant a
more complex simulation and show that our
stellar mass errors are accurate.

Several of our steps are only useful for putting the observable, the
stellar mass, in standard units and can be safely removed without affecting
the quality (intrinsic scatter) of the proposed mass proxy. 
We can skip the conversion from
luminosity to mass because this is a single number and therefore
has no effect on the scatter around the mean relation 
(it only changes the intercept, i.e.
the unit of the quantity being measured). 
In the same vein, we may also remove the correction for the missing flux,
which again only affects the intercept because it is a single number. 
Similarly, the conversion from
flux in a cylinder to flux into a sphere is a multiplicative term
with a negligible dependency on concentration (it then affects only 
the intercept). Because of the depth in the cluster rest-frame of
the photometry used, the light emitted by $r>19$ mag galaxies
is also negligible and could be ignored. In short, stellar mass
observationally measures the amount of light emitted from red detected 
galaxies, and several of our operations are only useful to obtain 
standard units for this quantity.  

As mentioned, these ``corrections" affect
the absolute value of the intercept of the mass-proxy relation. 
In other observational conditions (e.g. for shallow 
observations of distant clusters),
these corrections may take different values and differ
from cluster to cluster. Therefore, 
we prefer to characterise the proxy ``stellar mass" rather than 
the observation-dependent raw cluster luminosity, which likely 
has a larger scatter with mass (for widely varying observational data) and
an observation-dependent intercept. This choice also simplifies
the comparison with recent works.

\begin{table}
\caption{Stellar masses.}
\begin{tabular}{l l l}
\hline
ID & $\log_{10} M_{\star, r_{500}}$ & $\log_{10} M_{\star, r_{200}}$\\
  &  [$M_{\odot}$] & [$M_{\odot}$] \\
\hline
Abell ~160 & $ 12.33 \pm 0.12 $ & $ 12.48 \pm 0.10 $ \\ 
Abell 1177 & $ 12.05 \pm 0.15 $ & $ 12.10 \pm 0.16 $ \\ 
Abell 1795 & $ 12.66 \pm 0.05 $ & $ 12.80 \pm 0.05 $ \\ 
Abell 1991 & $ 12.46 \pm 0.12 $ & $ 12.64 \pm 0.09 $ \\ 
Abell 2029 & $ 12.59 \pm 0.08 $ & $ 12.70 \pm 0.08 $ \\ 
Abell 2092 & $ 12.34 \pm 0.10 $ & $ 12.44 \pm 0.08 $ \\ 
MKW 4 & $ 12.25 \pm 0.14 $  & $ 12.40 \pm 0.13 $ \\ 
NGC 4104 & $ 12.08 \pm 0.20 $& $ 12.12 \pm 0.25 $ \\ 
NGC 5098 & $ 12.11 \pm 0.15 $ & $ 12.29 \pm 0.12 $ \\ 
NGC 6269 & $ 12.39 \pm 0.12$  & $ 12.58 \pm 0.11 $ \\ 
RX J1022+3830 & $ 12.43 \pm 0.16$ & $ 12.51 \pm 0.11 $ \\ 
3C442A & $ 11.97 \pm 0.14 $&  $ 12.00 \pm 0.14 $ \\ 
\hline  
Abell 1991 & $ 12.48 \pm 0.11 $ & $ 12.66 \pm 0.09 $ \\ 
MKW 4  & $ 12.19 \pm 0.15 $ & $ 12.27 \pm 0.17 $ \\ 
\hline  
\end{tabular}
\hfill \break  
The last two lines indicate independently derived estimates based on
partially independent data. 
\end{table}

\section{Results}

Table 1 gives the stellar masses found  within $r_{500}$ and $r_{200}$ 
and their errors. Typically, stellar masses of clusters
in our sample have 0.11 dex errors. 
Two clusters appeared twice in our sample 
and were independently analysed based on
partially independent data (e.g. different background regions, partially 
different cluster regions because of the 
numerically different $r_{500}$ values, independent fits, 
different $c_{500}$).
We found nearly identical stellar masses 
showing the negligible effect of non-modelled sources of error.
Three clusters (Abell 160, MKW~4, and RXJ1022) are in common with
A10 and have nearly identical stellar masses within
$r_{200}$. A10 used a former release of the SDSS, which 
implements different
types of magnitudes, different $r_{200}$ values (derived using
a caustic analysis), and different background 
regions.

\begin{figure}
\centerline{%
\psfig{figure=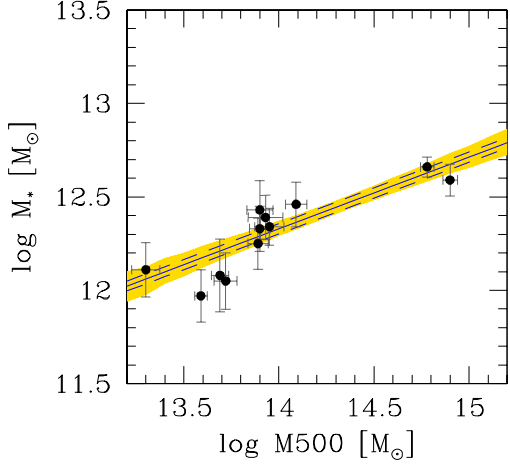,width=7truecm,clip=}
}
\caption[h]{Stellar vs total mass.
The mean relation between stellar mass and halo mass is marked with
a solid line, its 68\% uncertainty is shaded (in yellow). 
The dashed lines
show the mean relation plus or minus the (posterior mean) 
intrinsic scatter $\sigma_{scat}$. 
Error bars on the data points represent observed
errors for both variables. 
}
\label{fig:fig1}
\end{figure}

In order to fit the trend between stellar and total mass,
we use a standard Bayesian fitting model.  
Essentially, the model assumes that the true stellar mass 
and true halo mass are linearly related with some intrinsic scatter.
However, rather than having these true 
values, we have noisy measurements of both stellar mass and halo mass,
with noise amplitude different from point to point. The code 
to perform the computation is given 
in the Appendix A of A10 (ses also Andreon 2006, 2008; 
Andreon et al. 2006, 2008a, 2008b; Andreon \& Hurn 2010; Kelly 
2007).

Using the (fitting) model above, we found for our sample of 12 clusters:
\begin{eqnarray}
lgM_{\star, r_{500}} =  (0.38\pm0.07)\  (\log M_{500}-14.5) +12.53\pm0.04 \quad . \nonumber
\end{eqnarray}
Unless otherwise stated, results of the statistical computations 
are quoted in the form $x\pm y$, where $x$
is the posterior mean and $y$ is the posterior standard deviation.

\begin{figure}
\centerline{\psfig{figure=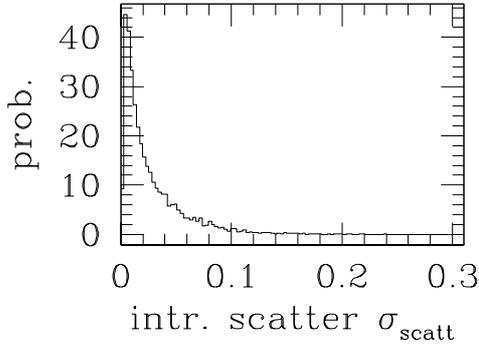,width=7truecm,clip=}}
\caption[h]{Posterior probability distribution of the intrinsic scatter
$\sigma_{lgM_{\star}|M_{500}}$.}
\end{figure}

\begin{figure}
\centerline{%
\psfig{figure=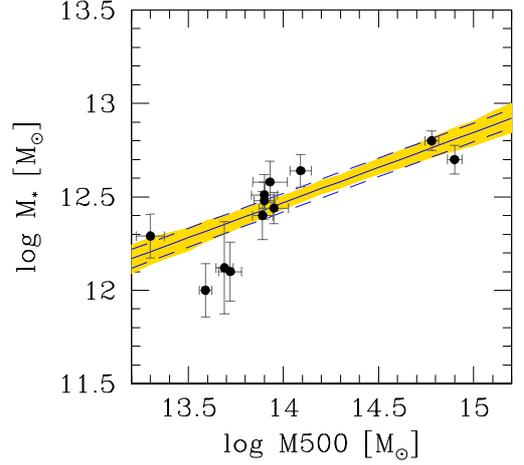,width=7truecm,clip=}
}
\caption[h]{Stellar vs total mass. As Fig.~1, but for stellar masses measured
within $r_{200}$.}
\label{fig:fig3}
\end{figure}

Figure~1 shows the relation between stellar mass and halo mass, observed
data, the mean scaling and its 68\% uncertainty 
and the mean intrinsic scatter around
the mean relation. The intercept and slope marginals
are well approximated by Gaussians. 
The (posterior) probability distribution of the intrinsic
scatter $\sigma_{scat}$ is an exponentially decreasing function (Fig.~2), 
i.e. the intrinsic
scatter is too small to be reliably measured with the data on hand.  The
same conclusion may also qualitatively inferred by noting that the mean model crosses 
the error bars of almost all points. 
Data are adequately described by the model: we generated 5000 fake
data sets from the model, computed their $\chi^2$, and found
that 71\% of them show a $\chi^2$ larger than the (true) data on hand
(an extreme p-value, say $\la 1$\% or $\ga 99$\% would call for a model
revision). In this p-value computation, we accounted for the uncertainty of the
regression parameters (intercept, slope, and intrinsic scatter)
by adopting Bayesian p-values (Gelman et al. 2004, 
Andreon 2011a).

The 90\% upper limit to the 
intrinsic stellar mass scatter at a given halo mass, 
$\sigma_{scat}=\sigma_{lgM_{\star}|M_{500}}$, is $0.06$ dex.
The mean and median values are $0.027$ and $0.017$ dex.

As mentioned in Sect. 2, we also measured stellar masses within $r_{200}$. 
Using the (fitting) model above, we found  
for our sample of 12 clusters:
\begin{eqnarray}
lgM_{\star, r_{200}} =  (0.38\pm0.07)\  (\log M_{500}-14.5) +12.65\pm0.05 \nonumber
\end{eqnarray}

Figure 3 shows the relation between stellar mass and halo mass, observed
data, the mean scaling and its 68\% uncertainty and the mean intrinsic 
scatter around the mean relation for these stellar masses. 

As for $r_{500}$, the intercept and slope marginals
are well approximated by Gaussians, and
the (posterior) probability distribution of the intrinsic
scatter $\sigma_{scat}$ is an exponentially decreasing function
with posterior mean $0.05\pm0.05$, median $0.03$, and 90\% upper limit of 
$0.12$ dex.
Also in this case, the data are adequately described by the model
(Bayesian p-value: 30\%).

Compared to the relation found using stellar masses measured within the smaller 
radius $r_{500}$, we found an identical slope, a larger
normalization (because stellar masses are integrated in larger apertures), 
and a somewhat looser constraint on the intrinsic scatter (expected
because we are now computing stellar masses in extrapolated $r_{200}$ 
values). Therefore, the tight scaling does not seem
to be a unique feature of the $r_{500}$ radius.

The slope, derived here from hydrostatic masses within $r_{500}$, 
is in agreement with what was found
in A10 for a much larger sample of 52 clusters using dynamical masses
within $r_{200}$. The latter sample
is not restricted to clusters appearing to be relaxed in X-ray
(unlike most samples dealing with $Y_X$) and consists
of a random sampling of an X-ray flux limited sample (as detailed
in A10 and also shown by a numerical simulation in Andreon \& Berg\'e 2012). 
A10 uses caustics and velocity dispersion-based
masses within $r_{200}$ and found a slope of $0.45\pm0.08$ and $0.53\pm0.08$
vs $0.38\pm0.07$ found in this work. The agreement of the
derived slope between hydrostatic-, caustics-, and 
velocity dispersion-based masses indicates the absence of a gross tilt
between the three mass scales. It also indicates that the current slope is not
driven by clusters at the extreme of the mass range. 
The A10 large sample also offers the opportunity of
checking if the found small $\sigma_{scatt}$ is  
an unwanted consequence of the small sample size.
To this aim, we recomputed the posterior distribution of $\sigma_{scatt}$
by forcing the slope to be $0.45\pm0.08$ (i.e. adopting
the latter as prior, in place of the originally adopted Student-$t$ prior)
and found an indistinguishable posterior probability distribution. This
indicates that the small scatter is not due to overfitting.

The scatter we found in the present work is lower than, although consistent 
with at 95\%, the scatter found in A10 using
noisier masses ($0.14$ vs $0.05$ dex errors for masses within $r_{200}$). As mentioned, intrinsic
scatter values smaller than the mass errors (like those in A10) should
be read as upper limits, hence increasing the agreement. 
Indeed, A10 assumed Gaussian errors (likelihood
function) for mass, whereas the later work by Serra et al. (2011) found
errors to be asymmetric (i.e. the likelihood is not Gaussian). 
If the uncertainty in the likelihood shape and in the 
error noisiness were accounted for, a larger uncertainty
on the intrinsic scatter would have been found by A10, improving the current
(95\%) agreement. 

More in general, imprecise measurements of
mass (i.e. with $\gg 0.05$ dex errors) are the probable reason
why the tight stellar mass vs halo mass has not been noted before: imprecise
errors not accounted for in the statistical analysis tend
to leave a residual scatter of the order of a fraction of the mass error,
which the fitting algorithm attributes to the intrinsic scatter term.

\subsection{The usefulness of stellar masses for cluster samples}

How does the scatter in stellar mass at a given halo mass 
compare to the scatter of other mass proxies?  
In this comparison, we strictly follow
Sect. 3.3.4 of the  Allen et al. (2011) review paper.

- the 90\% upper limit of the scatter of $Y_X$ at a given mass
is $0.06$ dex (Mantz et al. 2010a) above $lgM \sim 14.5$ $M_\odot$ 
(vs $0.06$ dex for stellar mass within $r_{500}$). 
Other works (Arnaud et al. 2007,
Kravtsov et al. 2006) report compatible point estimates, but
with unspecified uncertainties. 

- the upper limit of Mantz et al. (2010a)  to 
the scatter of the core-excised X-ray luminosity at a given mass
is $\la 0.04$ dex (at an unspecified level) vs $0.06$ dex for stellar mass
(90\% upper limit).

To summarise, the stellar mass is a mass proxy
with a scatter similar to best-mass proxies.
Furthermore,
the range in mass in which the relation holds extends to lower masses
than other proxies, which is advantageous
from a cosmological perspective because clusters of lower mass 
are far more abundant than more massive clusters. On the other
hand, one should not jump to conclusions:
the relative merit of different mass proxies cannot be
estimated by any single number (e.g. intrinsic scatter).  
One should also consider the
slope of the mass-proxy relation  (stellar mass has a shallower
slope than core-excised X-ray luminosity 
and $Y_X$) and the size of the sample for which the mass proxy
can be measured (stellar mass has probably the largest sample size,
being measurable from survey data, see Sect. 3.3), etc.
Nevertheless, one of the parameters in cosmological
estimates using galaxy cluster counts and using stellar
mass as mass proxy is now known and found to be tiny at most.

Our sample size, 12 clusters, is small. However,
Vikhlinin et al. (2009b) measured 
the scatter with only four additional clusters, 
the widely cited small $Y_X$ scatter of Vikhlinin et al. (2006)
and Arnaud et al. (2007) is derived analysing smaller
samples, and
Kravtsov et al. (2006) proposed the $Y_X$ mass proxy with only four more 
(simulated) clusters. We have already commented that the scatter remains
small even forcing the slope to be the one derived
from a much larger cluster sample (with noisier mass errors).
Mantz et al. (2010a,b) have a much larger sample, but their $M_{500}$ are
obtained via a complex and indirect path passing
through $f_{gas}$ (see their Sect. 2.2.3).

Similarly, concerns 
that the scatter is found small because of overestimated halo
mass errors should be allayed by the fact 
that: a) there is some overlap among cluster samples; four
of our clusters are in common with Vikhlinin et al. (2006), three with
Vikhlinin et al. (2009b, they removed MKW4) 
and that all halo mass errors, for common and uncommon clusters, 
have been derived by a single team in the same way; and that b) 
even assuming that mass errors are zero (an unbelievably small value)
in order to maximally boost the intrinsic scatter,
its posterior is almost unchanged, i.e. stellar mass compares 
as favorably as the best-mass proxies even in this 
implausible case.  Finally, our numerical simulation in Sect. 2 
shows that our stellar mass errors are accurate, and thus
we exclude that the small $\sigma_{scatt} $ are due
to inaccurate (overestimated) errors on
stellar mass.

Two comments about cluster physics are in order.
First, a small scatter between stellar mass in red galaxies
and halo mass is expected if during cluster build-up at most a small fraction 
of galaxies is added to the existing red population, i.e. if
the fraction of galaxies that turns from blue to red is small. This is
likely the case because the fraction (in number) of blue galaxies is small
in clusters at all redshift (Raichoor \& Andreon 2012; Andreon et al. 2006).
Moreover, blue galaxies have, on average, lower masses than red ones, and thus
their contribution to the total stellar mass is negligible
(e.g. Fukugita et al. 1998; Girardi et al. 2000).
The found small scatter is instead at variance with numerical simulations (e.g. Young
et al. 2011). However, these fail to reproduce, by a large factor, both
the stellar fraction and the amount of intracluster light, indicating
that probably sub-grid physics is not yet accurately known/implemented.
 
Second, since the intrinsic scatter of the stellar mass fraction is
negligible (all clusters have the same stellar mass at a given
cluster mass), the known observed intrinsic scatter in the gas fraction (Sun et
al. 2009, A10) cannot be compensated by a  correlated scatter in stellar mass
fraction to keep constant the total baryon fraction.

\subsection{The usefulness of stellar masses for individual clusters}

In the previous section, we computed the scatter in observable at a given mass,
$\sigma_{proxy|M}$.  
As mentioned in the introduction, cosmological estimates need $p(proxy|M)$, whose
minimal parametrisation is given by a Gaussian with $\sigma=\sigma_{proxy|M}$ 
(see, e.g. Allen et al. 2011 or Weinberg et al. 2012
reviews).  In fact, Vikhlinin et
al. (2009), Mantz et al. (2010b), and Rozo et al. (2010) used 
$\sigma_{proxy|M}$ in their cosmological estimates using cluster counts.
For the stellar mass, we found in the previous section a 90\% upper limit of
$0.06$ dex on $\sigma_{proxy|M}$.

The inverse scatter, $\sigma_{M|proxy}$, is also interesting. As mentioned
in the introduction, it is the minimal error of the estimated mass of a given
cluster. As detailed in Andreon \& Hurn (2010), the error in the proxy itself, and 
of course, its availability, limit the usefulness of a given proxy.
We computed the inverse scatter
as in previous section, fitting the same data set, but using
stellar mass as the predictor variable. We found
\begin{eqnarray}
\log M_{500} =  (2.02\pm0.35)\  (lgM_{\star, r_{500}} -12.5) +14.39\pm0.09 \quad . \nonumber
\end{eqnarray}
The posterior probability distribution of the 
intrinsic scatter $\sigma_{lg M_{500}| M_\star}$ is an exponentially decreasing
function, indicating that the scatter is too small to be determined with 
the data on hand, with mean $0.08$ dex, median $0.04$ dex, and
90\% upper limit $0.2$ dex.
These numbers quantify the uncertainty of the mass of a single 
cluster predicted from its $lgM_{\star, r_{500}}$, i.e. its optical 
luminosity, if all other
sources of errors (the above proxy-mass calibration and the
proxy uncertainty) are negligible. However, this is unlikely to be
the case of stellar masses for clusters with $lgM\la 14$ $M_\odot$, 
which are also the
majority in our sample. This is because they have a typical $0.1$ dex error in
stellar mass
(Table 1), which propagates into a $0.2$ dex error in the predicted mass. 
The posterior predictive mass uncertainty, which 
encapsulates both proxy errors and errors coming from the proxy-mass calibration,
turns out to be, on average, $0.32$ dex for these not massive systems. For our
two most massive systems, the predictive mass uncertainty is $0.22$ and $0.25$ dex,
whereas for the very rich MS1054 cluster discussed in next section it is
$0.21$ dex.

These uncertainties are larger than those one may obtain
from high-quality 
X-ray-based mass estimates, such as $Y_X$ or core-excised X-ray 
luminosities. 
We emphasise, however, that the value of a mass proxy also depends on
its availability and that obtaining precise estimates of
$Y_X$ or core-excised X-ray luminosities
requires expensive, even unfesable, follow-up observations from space. 
The value of a 
mass proxy also depends on how well the mass-proxy relation is calibrated.
The $Y_X$ and core-excised X--ray luminosity vs 
mass relations are un-calibrated below $lgM=14-14.5$ 
$M_\odot$, the precise value depends on which mass proxy is considered.
This is a mass quite large mass value in the cluster mass function 
and even more so at intermediate and high redshift.
Therefore, stellar masses are a useful mass proxy for clusters of not large mass 
or when one may not afford the luxury of having the needed
high--quality follow--up observations to derive more precise 
predicted masses. 

Strategies to reduce the size of stellar mass errors and
thus to increase its quality as mass estimator for individual
clusters, are being investigated. 

\begin{figure}
\centerline{%
\psfig{figure=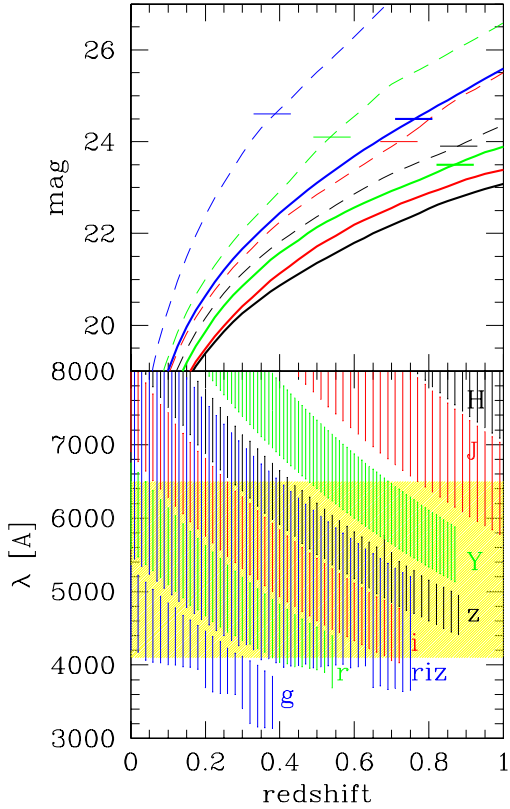,width=7truecm,clip=}
}
\caption[h]{Depth and wavelength coverage of DES and Eclid surveys. {\it Upper panel:}
$g$,$r$,$i$, and $z$ (from left to right) filters are indicated with thin
dashed (blue, green, red and black) lines. 
$riz$, $y$, $J$, and $H$ (from left to right) Eclid filters are indicated with 
thick solid (blue, green,
red, and black) lines. The horizontal tick indicates the $\sim 10 \sigma$ depth.
{\it Bottom panel:} Wavelength coverage of the filters for redshift bins, where galaxies are brighter 
than the $\sim10 \sigma$ depth. The shaded (yellow) region marks the wavelength sampling
of $g-r$ at $z<0.08$.}
\label{fig:fig3}
\end{figure}

\subsection{On which part of the Universe is stellar mass measurable with current data?}

The analysis of previous sections uses $g-r$ colour and luminosities
of galaxies brighter than $r<19$ in the nearby Universe ($z<0.08$). 
Figure 4 illustrates how these constraints change with redshifts.
The top panel illustrates the apparent luminosity of a red $z=0.08$ $r=19$ galaxy,
modelled as a $z_f=5$ single stellar population using the 2007 version of the Bruzual
\& Charlot (2003) synthesis population model, for different filters: $g$, $r$,
$i$, and $z$ for the Dark Energy Survey
(Abbott et al. 2005) 
and $riz$, $Y$, $J$, and $H$ for Euclid
(Laureijs et al. 2011) with the corresponding $\sim10 \sigma$ depth (horizontal ticks). 
The bottom
panel illustrates the $\lambda$ range sampled by these filters. Only 
redshift bins where galaxies are brighter than the 10 $\sigma$ depth are plotted. 
The shaded yellow is the $\lambda$ range sampled by $g-r$ at $z<0.08$. As the figure
shows,
we always have two filters in the shaded region, i.e. up to $z=0.9$,
these data have the depth and wavelength coverage appropriate to measure
stellar masses as we did at $z<0.08$. 
Indeed, an even larger redshift range is accessible for the massive
clusters, because the depth requirement can be safely relaxed, as it is not
necessary 
to accurately measure stellar masses for these objects. We further emphasise
that these depths, which cover
several thousands of deg$^2$,
are quite comparable to the depths of surveys available right now 
covering a ten of deg$^2$,
such as the deep fields of the CFHTLS (Cuillandre \& Bertin 2006),  VISTA-VIDEO\footnote{
http://star-www.herts.ac.uk/\%7Emjarvis/video/index.html} (Jarvis et al. 2012),
or WIRDS (Bielby et al. 2011). Repeating this exercise with the planned 
Large Sinoptic Survey Telescope
10 yr integration (Ivezic et al. 2008), we found that 
$z=1.2$ may be easily reached.

In order to use stellar masses for large cluster samples, the integral
of the luminosity function computation should not include any manual step
or complementary (e.g. spectroscopic) data.
In our analysis of nearby clusters, we removed by hand 
saturated stars misclassified as galaxies and used spectroscopy
to identify foreground galaxies brighter and larger than the BCG. 
The first step can be automatically implemented by
using the central object intensity (or any flag based on it). The second step
may use galaxy sizes and photometric redshifts in place of
spectroscopic redshifts that are perhaps not available. Therefore, automatic
computation is possible for large cluster samples.

In order to test whether 
the increased galaxy background at high redshift may be detrimental for an
accurate measurement of the stellar mass, we computed the stellar mass
of two high redshift clusters using real data.

\begin{figure}
\centerline{\psfig{figure=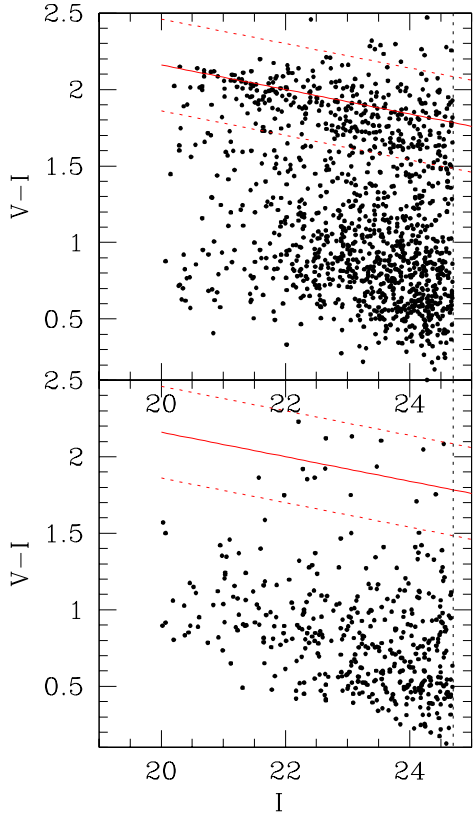,width=7truecm,clip=}}
\caption[h]{Colour--magnitude relation in the cluster (top panel) 
and control field (bottom panel) line-of-sight directions. The
solid line marks the colour-magnitude relation, whereas the dotted
lines delimit the portion of plane that qualifies galaxies to
be called red. The vertical line marks the considered limiting depth. 
In this figure, we use magnitudes in the Vega system, for comparison
with the similar figures in Andreon (2006).
}
\end{figure}

We first consider MS1054 (Gioia et al. 1990) at $z=0.83$, a 
massive cluster of galaxies ($lgM_{500} \approx 14.8$ $M_\odot$, Jee et al. 2005) for which 
we consider only $V<24.7$ mag (to match depths) and $V-I$ colour, 
coming from FIRES (Labbe et al. 2003; F\"orster Schreiber et al. 
2006), which was already used in Andreon (2006) to measure the luminosity function
of its red galaxies (see his Figure 4). 
Figure 5 shows the $V-I$ vs $I$ diagram of galaxies in the cluster
line-of-sight (top panel) and in a reference line-of-sight (bottom panel). 
As is fairly obvious, the background contribution,
depicted in the bottom panel, at the colour of red galaxies (inside the dashed
lines), is negligible.  
Our luminosity function determination of red galaxies
of several clusters (e.g. of the 28 clusters
in Andreon 2008, up to $z=1.3$) observationally confirms 
that a low background can be easily achieved by adopting a colour index
bracketing the 4000 $\AA$ \  break and that the background
is not a large penalty for red galaxies up to the highest redshifts. Furthermore, the
background contamination can be further reduced using photometric redshift. 
This situation extends up to the largest redshifts: 
Andreon (2011b) found that there are
on average 3.6 back/foreground red galaxies in the $z\sim2.2$ 
JKCS041 cluster (Andreon et al. 2009, Andreon \& Huertas-Company 2011)
line-of-sight from measurements
all around the cluster vs 23
galaxies on the red sequence.
We computed the MS1054 stellar mass as we did for lower redshift clusters and 
found an error of $0.04$ dex. The error is small because of the large richness
of MS1054 and the low background contamination at the red sequence colour.

As a worse case scenario, we consider RzCS052 at $z=1.016$ 
(Andreon et al. 2008a,b). The cluster is still a massive one
($lgM_{500} \approx 14.4$ $M_\odot$, Andreon et al. 2008a), barely detected
in $\sim 10$ ks XMM observations\footnote{ 
There is a typo into the exponent of the cluster flux reported in Andreon et al. 
2008a: the correct flux is $1.2 \ 10^{-14}$ erg$^{-1}$ s$^{-1}$ cm$^{-2}$, 
as one may discern from the quoted $L_X$.}.
The cluster is at a redshift larger than the maximal one that we conservatively 
said possible
for stellar mass computation, and even more so with the available data 
(from Andreon et al. 2008b), which are shallower (by a few tenths of mag)
than the DES+Euclid surveys, and made the derivation of stellar mass challenging.
The colour-magnitude
plot and luminosity function of the red galaxies of RzCS052
are depicted in Fig 2 and 4 of Andreon et al. (2008b).
We computed the stellar mass as we did for lower redshift clusters and 
found an error of $0.1$ dex, in spite of having taken
a cluster outside the redshift range claimed to be accessible, having
used data of lower depth than upcoming surveys, and having discarded the multicolour
data available for this cluster (and for the considered surveys).

As for other mass proxies (e.g. $Y_X$ and $L_X$), 
the quoted error assumes to perfectly know the evolution
of the intercept of the relation between mass and mass proxy.
In the case of stellar masses, the evolution is modulated by the M/L ratio 
of red galaxies, which is robustly known (e.g. Treu et al.
2005, Holden et al. 2010) up to at least $z\sim1$. Instead, 
the evolution of
other mass proxies is poorly known (e.g. the evolution of $Y_X$ or 
$L_X$ with $z$, Andreon, Trinchieri \& Pizzolato 2011).

\section{Summary}

The analysis of a sample of clusters of the nearby Universe ($z<0.08$), which is
of low cardinality because of the need of high-quality mass estimates ($0.05$ dex 
mass errors), shows that stellar mass, derived measuring 
the luminosity emitted by red galaxies only, 
has a tiny scatter with
halo mass, similar to best-mass proxies. We show that stellar masses can
be measured with survey-quality data (optical and near-infrared imaging)
typical of current (on some square degrees) and upcoming 
(on several thousand of square degrees) surveys up to $z=0.9-1.2$. 

Three-quarters of 
the studied clusters have $lgM\la 14$ $M_\odot$, which is advantageous
from a cosmological perspective because these clusters are far more
abundant than more massive clusters. In constrast to other mass
proxies, we
have robust knowledge about the evolution of the stellar mass proxy at a 
given halo mass. This is because the mass evolution is modulated by 
the evolution of the 
M/L of red galaxies, which is observationally well constrained 
up to at least $z \sim 1$.  We emphasise, however, that a small
scatter is an essential property for a good mass proxy, but this number 
alone cannot completely characterise a mass proxy: availability of
the mass proxy and its error also are two essential properties. We also emphasise
that if the purpose of a mass proxy is to derive the mass of an
individual cluster (as opposed to a sample), 
survey-based stellar masses should be viewed as a
complementary mass proxy to be used
when one may not afford the luxury of having  
non-survey high-quality masses requested for better mass proxies,
such as $Y_X$ or core-excised luminosities, or when these measurements are 
infeasible or uncalibrated.

Clusters in our sample display an intrinsic scatter in gas fraction
(Sun et al. 2009, Andreon 2010) and a tiny scatter in stellar
fraction. Therefore, the latter cannot compensate for the scatter of the former
to keep the total baryon fraction constant. 

\begin{acknowledgements} 
I thank Antonaldo Diaferio, Fabio Gastaldello and Federico
Marulli for useful discussions. I warmly thank Ben Maughan for
useful suggestions that helped to improve the quality of my paper.
For the standard SDSS, NED, and DSS acknowledgements see: 
http://www.sdss3.org/, 
http://nedwww.ipac.caltech.edu/, and
http://archive.stsci.edu/dss/acknowledging.html.
\end{acknowledgements}


\begin{thebibliography}{}

\bibitem[The Dark Energy Survey Collaboration(2005)]{2005astro.ph.10346T} 
Abbott et al., 2005, The Dark Energy Survey (arXiv:astro-ph/0510346) 

\bibitem[Aihara et al.(2011)]{2011ApJS..193...29A} 
Aihara, H., Allende  Prieto, C., An, D., et al.\ 2011, ApJS, 193, 29 

\bibitem[Allen et al.(2011)]{2011ARA&A..49..409A} 
Allen, S.~W., Evrard, A.~E., \& Mantz, A.~B.\ 2011, ARAA, 49, 409 

\bibitem[Andreon(2002)]{2002A&A...382..495A} 
Andreon, S.\ 2002, A\&A, 382, 495 

\bibitem[Andreon(2006)]{2006MNRAS.369..969A} 
Andreon, S.\ 2006, MNRAS, 369, 969 

\bibitem[\protect\citeauthoryear{Andreon}{2008}]{2008MNRAS.386.1045A} 
Andreon S., 2008, MNRAS, 386, 1045 

\bibitem[Andreon(2010)]{2010MNRAS.407..263A} 
Andreon, S.\ 2010, MNRAS, 407, 263 (A10)

\bibitem[Andreon(2011)]{2011arXiv1112.3652A} 
Andreon, S.\ 2011a, in Astrostatistical Challenges for the New Astronomy,
ed. J. Hilbe, Springer Series on Astrostatistics (arXiv:1112.3652) 

\bibitem[Andreon(2011)]{2011A&A...529L...5A} 
Andreon, S.\ 2011b, A\&A, 529, L5 

\bibitem[Andreon  \& Berg\'e(2012)]{xxx} 
Andreon, S., \& Berg\'e, J.\ 2012, A\&A, in press (arXiv:1209.5938)

\bibitem[Andreon \& Huertas-Company(2011)]{2011A&A...526A..11A} 
Andreon, S., \& Huertas-Company, M.\ 2011, A\&A, 526, A11 

\bibitem[Andreon  \& Hurn(2010)]{2010MNRAS.404.1922A} 
Andreon, S., \& Hurn, M.~A.\ 2010, MNRAS, 404, 1922 

\bibitem[]{} 
Andreon, S., Punzi, G., Grado, A., 2005, MNRAS, 360, 727

\bibitem[Andreon et al.(2006)]{2006MNRAS.372...60A} 
Andreon, S., Cuillandre, J.-C., Puddu, E., \& Mellier, Y.\ 2006, 
        MNRAS, 372, 60

\bibitem[Andreon et al.(2006)]{2006MNRAS.365..915A} 
Andreon, S., Quintana, H., Tajer, M., Galaz, G., \& Surdej, J.\ 2006, 
	MNRAS, 365, 915

\bibitem[Andreon et al.(2008)]{2008MNRAS.385..979A} 
Andreon, S., Puddu, E., de Propris, R., \& Cuillandre, J.-C.\ 2008b, 
        MNRAS, 385, 979 

\bibitem[Andreon et al.(2008)]{2008MNRAS.383..102A} 
Andreon, S., De Propris, R., Puddu, E., Giordano, L., \& Quintana, 
        H.\ 2008a, MNRAS, 383, 102 

\bibitem[Andreon et  al.(2009)]{2009A&A...507..147A} 
Andreon, S., Maughan, B., Trinchieri, G., \& Kurk, J.\ 2009, A\&A, 507, 147 


\bibitem[Andreon et al.(2011)]{2011MNRAS.412.2391A} 
Andreon, S., Trinchieri, G., \& Pizzolato, F.\ 2011, MNRAS, 412, 2391 

\bibitem[Arnaud et al.(2007)]{2007A&A...474L..37A} 
Arnaud, M., Pointecouteau, E., \& Pratt, G.~W.\ 2007, A\&A, 474, L37 

\bibitem[Bielby et al.(2011)]{2011arXiv1111.6997B} 
Bielby, R., Hudelot, P., McCracken, H.~J., et al.\ 2011, A\&A, submitted
	(arXiv:1111.6997) 

\bibitem[Blanton et al.(2001)]{2001AJ....121.2358B} 
Blanton, M.~R., et al.\ 2001, AJ, 121, 2358 

\bibitem[Bruzual \& Charlot(2003)]{2003MNRAS.344.1000B} 
Bruzual, G., \& Charlot, S.\ 2003, MNRAS, 344, 1000 

\bibitem[Cappellari et al.(2006)]{2006MNRAS.366.1126C} 
Cappellari, M., et  al.\ 2006, MNRAS, 366, 1126 

\bibitem[Cuillandre \& Bertin(2006)]{2006sf2a.conf..265C} 
Cuillandre, J.-C., \& Bertin, E.\ 2006, SF2A-2006: Semaine de l'Astrophysique 
	Francaise, 265 

\bibitem[Dai et al.(2010)]{2010ApJ...719..119D} 
Dai, X., Bregman, J.~N., Kochanek, C.~S., \& Rasia, E.\ 2010, ApJ, 719, 119 

\bibitem[D'Agostini(2011)]{2011arXiv1112.3620D} 
D'Agostini, G.\ 2011, unpublished (arXiv:1112.3620) 


\bibitem[F{\"o}rster Schreiber et al.(2006)]{2006AJ....131.1891F} 
F{\"o}rster Schreiber, N.~M., Franx, M., Labb{\'e}, I., et al.\ 2006, AJ, 131, 1891 

\bibitem[\protect\citeauthoryear{Fukugita, Hogan, \& Peebles}{1998}]{1998ApJ...503..518F} 
Fukugita M., Hogan C.~J., Peebles P.~J.~E., 1998, ApJ, 503, 518 


\bibitem[]{Gel04}
Gelman A., Carlin J., Stern H., Rubin D., 2004,
``Bayesian Data Analysis", (Chapman \& Hall/CRC)

\bibitem[Gioia et al.(1990)]{1990ApJS...72..567G} 
Gioia, I.~M., Maccacaro, T., Schild, R.~E., Wolter, A., Stocke, J.~T., Morris, S.~L., \& Henry, 
J.~P.\ 1990, ApJS, 72, 567 

\bibitem[\protect\citeauthoryear{Girardi et al.}{2000}]{2000ApJ...530...62G} 
Girardi M., Borgani S., Giuricin G., Mardirossian F., Mezzetti M., 2000, ApJ, 530, 62 

\bibitem[\protect\citeauthoryear{Gonzalez, Zaritsky, \& Zabludoff}{2007}]{2007ApJ...666..147G} 
Gonzalez A.~H., Zaritsky D., Zabludoff A.~I., 2007, ApJ, 666, 147 

\bibitem[Holden et al.(2010)]{2010ApJ...724..714H} 
Holden, B.~P., van der Wel, A., Kelson, D.~D., Franx, M., 
	\& Illingworth, G.~D.\ 2010, ApJ, 724, 714 

\bibitem[Humphrey et al.(2012)]{2012ApJ...755..166H} 
Humphrey, P.~J., Buote, D.~A., O'Sullivan, E., \& Ponman, T.~J.\ 2012, ApJ, 755, 166 


\bibitem[Humphrey et al.(2011)]{2011ApJ...729...53H} 
Humphrey, P.~J., Buote, D.~A., Canizares, C.~R., Fabian, A.~C., 
		\& Miller, J.~M.\ 2011, ApJ, 729, 53 

\bibitem[Ivezi{\'c} et al.(2002)]{2002AJ....124.2364I} 
Ivezi{\'c}, {\v Z}., et al.\ 2002, AJ, 124, 2364 

\bibitem[Ivezic et al.(2008)]{2008arXiv0805.2366I} 
Ivezic, Z., et al.\ 2008, unpublished, arXiv:0805.2366 

\bibitem[Jarvis et al.(2012)]{2012arXiv1206.4263J} 
Jarvis, M.~J., Bonfield, D.~G., Bruce, V.~A., et al.\ 2012, MNRAS, in press
	(arXiv:1206.4263) 

\bibitem[Jee et al.(2005)]{2005ApJ...634..813J} 
Jee, M.~J., White, R.~L., Ford, H.~C., Blakeslee, J.~P., Illingworth, G.~D., Coe, D.~A., \& Tran, 
K.-V.~H.\ 2005, ApJ, 634, 813 

\bibitem[Jenkins et al.(2001)]{2001MNRAS.321..372J} 
Jenkins, A., Frenk, C.~S., White, S.~D.~M., et al.\ 2001, MNRAS, 321, 372 

\bibitem[Kelly(2007)]{2007ApJ...665.1489K} 
Kelly, B.~C.\ 2007, ApJ, 665, 1489 

\bibitem[Kravtsov et al.(2006)]{2006ApJ...650..128K} 
Kravtsov, A.~V., Vikhlinin, A., \& Nagai, D.\ 2006, ApJ, 650, 128 

\bibitem[Labb{\'e} et al.(2003)]{2003AJ....125.1107L} 
Labb{\'e}, I., et al.\ 2003, AJ, 125, 1107 

\bibitem[Laureijs et al.(2011)]{2011arXiv1110.3193L} 
Laureijs, R., Amiaux, J., Arduini, S., et al.\ 2011, Euclid Definition Study Report
(arXiv:1110.3193) 

\bibitem[Lima \& Hu(2005)]{2005PhRvD..72d3006L} 
Lima, M., \& Hu, W.\ 2005,  Phys. Rev. D, 72, 043006 

\bibitem[Lupton et al.(2002)]{2002SPIE.4836..350L} 
Lupton, R.~H., Ivezic, Z., Gunn, J.~E., Knapp, G., Strauss, M.~A., 
	\& Yasuda, N.\ 2002, Proc Spiee, 4836, 350

\bibitem[Mantz et al.(2010)]{2010MNRAS.406.1773M} 
Mantz, A., Allen, S.~W., Ebeling, H., Rapetti, D., \& Drlica-Wagner, A.\ 2010a, 
MNRAS, 406, 1773

\bibitem[Mantz et al.(2010)]{2010MNRAS.406.1759M} 
Mantz, A., Allen, S.~W., Rapetti, D., \& Ebeling, H.\ 2010b, MNRAS, 406, 1759 

\bibitem[Meyers et al.(2012)]{2012ApJ...750....1M} 
Meyers, J., Aldering, G., Barbary, K., et al.\ 2012, ApJ, 750, 1 

\bibitem[Navarro et al.(1997)]{1997ApJ...490..493N} 
Navarro, J.~F., Frenk, C.~S., \& White, S.~D.~M.\ 1997, ApJ, 490, 493 

\bibitem[Oemler(1974)]{1974ApJ...194....1O} 
Oemler, A.~J.\ 1974, ApJ, 194, 1

\bibitem[Press \& Schechter(1974)]{1974ApJ...187..425P} 
Press, W.~H., \& Schechter, P.\ 1974, ApJ, 187, 425 

\bibitem[Randall et al.(2009)]{2009ApJ...700.1404R} 
Randall, S.~W., Jones, C., Markevitch, M., et al.\ 2009, ApJ, 700, 1404 

\bibitem[Raichoor \& Andreon(2012)]{2012A&A...543A..19R} 
Raichoor, A., \& Andreon, S.\ 2012, A\&A, 543, A19 

\bibitem[Rozo et al.(2010)]{2010ApJ...708..645R} 
Rozo, E., Wechsler, R.~H., Rykoff, E.~S., et al.\ 2010, ApJ, 708, 645 

\bibitem[Sandage, Tammann, \& Yahil(1979)]{1979ApJ...232..352S} 
Sandage,  A., Tammann, G.~A., \& Yahil, A.\ 1979, ApJ, 232, 352 

\bibitem[Schechter(1976)]{1976ApJ...203..297S} 
Schechter, P.\ 1976, ApJ, 203, 297

\bibitem[Serra et al.(2011)]{2011MNRAS.412..800S} 
Serra, A.~L., Diaferio, A., Murante, G., \& Borgani, S.\ 2011, MNRAS, 412, 800 

\bibitem[\protect\citeauthoryear{Sun et al.}{2009}]{2009ApJ...693.1142S} 
Sun M., Voit G.~M., Donahue M., Jones C., Forman W., Vikhlinin A., 2009, 
	ApJ, 693, 1142 

\bibitem[\protect\citeauthoryear{Tremaine \& Richstone}{1977}]{1977ApJ...212..311T} 
Tremaine S.~D., Richstone D.~O., 1977, ApJ, 212, 311 

\bibitem[Treu et al.(2005)]{2005ApJ...622L...5T} 
Treu, T., Ellis, R.~S., Liao, T.~X., \& van Dokkum, P.~G.\ 2005, ApJ, 622, L5 

\bibitem[Vikhlinin et al.(2006)]{2006ApJ...640..691V} 
Vikhlinin, A., Kravtsov, A., Forman, W., Jones, C., Markevitch, 
	M., Murray, S.~S., \& Van Speybroeck, L.\ 2006, ApJ, 640, 691 

\bibitem[Vikhlinin et al.(2009)]{2009ApJ...692.1033V} 
Vikhlinin, A., Burenin, R.~A., Ebeling, H., et al.\ 2009, ApJ, 692, 1033 

\bibitem[Vikhlinin et al.(2009)]{2009ApJ...692.1060V} 
Vikhlinin, A., Kravtsov, A.~V., Burenin, R.~A., et al.\ 2009, ApJ, 692, 1060 


\bibitem[Young et al.(2011)]{2011MNRAS.413..691Y} 
Young, O.~E., Thomas, P.~A., Short, C.~J., \& Pearce, F.\ 2011, MNRAS, 413, 691 

\bibitem[Weinberg et al.(2012)]{2012arXiv1201.2434W} 
Weinberg, D.~H., Mortonson, M.~J., Eisenstein, D.~J., et al.\ 2012, 
	Physics Reports, in press (arXiv:1201.2434) 

\bibitem[Zwicky(1957)]{1957moas.book.....Z} 
Zwicky, F.\ 1957, Morphological astronomy, Berlin: Springer


\end{thebibliography}
\end{document}